\documentclass[%
 reprint,
showpacs,preprintnumbers,
 amsmath,amssymb,
 aps,
 pra,
floatfix,
]{revtex4-1}

\usepackage{graphicx}
\usepackage{dcolumn}
\usepackage{bm}
\usepackage{hyperref}

\begin{document}


\title{Microwave spectroscopy of the $1\mathrm{s}n\mathrm{p}\,^3\mathrm{P}_J$ fine structure of high Rydberg states in $^4$He}

\author{A. Deller}
\email{a.deller@ucl.ac.uk}
\author{S. D. Hogan}
\email{s.hogan@ucl.ac.uk}
\affiliation{Department of Physics and Astronomy, University College London, Gower Street, London WC1E 6BT, U.K.}%

\date{\today}

\begin{abstract}
The $1\mathrm{s}n\mathrm{p}\,^3\mathrm{P}_J$ fine structure of high Rydberg states in helium has been measured by microwave spectroscopy of single-photon transitions from $1\mathrm{s}n\mathrm{s}\,^3\mathrm{S}_1$ levels in pulsed supersonic beams. For states with principal quantum numbers in the range from $n=34$ to 36, the $J = 0 \rightarrow 2$ and $J = 1 \rightarrow 2$ fine structure intervals were both observed.  For values of $n$ between~45 and~51 only the larger $J = 0 \rightarrow 2$ interval was resolved. The experimental results are in good agreement with theoretical predictions. Detailed characterization of residual uncancelled electric and magnetic fields in the experimental apparatus, and calculations of the Stark and Zeeman structures of the Rydberg states in weak fields, were used to quantify systematic contributions to the uncertainties in the measurements. 
\end{abstract}

\maketitle


\section{\label{sec:intro}Introduction}
High-resolution spectroscopy of the energy-level structure of simple atomic systems containing two leptons -- such as the helium atom, positronium atom, or hydrogen molecule -- plays an important role in experimental tests of bound-state quantum electrodynamics (QED)~\cite{karshenboim2005,sprecher2011}. In the case of the helium atom, this has motivated a wide range of studies from the vacuum-ultraviolet to the infrared and microwave regions of the electromagnetic spectrum~\cite{herzberg58,baig84,eikema93,bergeson98,schulz96,vanRooij11}.  In particular, since being identified as a sensitive probe of the fine-structure constant, $\alpha$,~\cite{Schwartz1964} the fine structure of the $1\mathrm{s}2\mathrm{p}\,^3\mathrm{P}$ term~\cite{Houston1927} has attracted significant theoretical and experimental interest  \cite[e.g.,][]{george2001, Drake2002, Giusfredi2005, Pachucki2006, Zelevinsky2005, Zheng2017}. Similarly, precise measurements and calculations of the $1\mathrm{s}n\mathrm{p}\,^3\mathrm{P}_J$ fine structure, for principal quantum numbers $n \leq 10$, have also been performed \cite{Mueller2005, hessels1992,claytor1995,storry1995,stevens1999}.

For higher values of $n$, the fine structure intervals and the magnitude of the corresponding QED corrections reduce significantly. However, precise measurements of the fine structure of the triplet Rydberg states are essential for a number of recently developed experiments. These include experiments that involve coupling Rydberg helium atoms to chip-based microwave circuits for applications in hybrid cavity quantum electrodynamics and quantum information processing~\cite{hogan2012,thiele2015,rabl2006}, and studies of F\"orster resonance energy transfer in collisions with polar ground-state molecules~\cite{zhelyazkova2017}. In these cases, Rydberg states with $n = 30$ to $70$ and values of the electron orbital angular momentum quantum number $0 \leq \ell \leq (n - 1)$~\cite{zhelyazkova2016} are employed. In general these experiments take advantage of: (i) the minimal detrimental effects caused by helium adsorption within the experimental apparatus, particularly in cryogenic environments~\cite{hatterman2012,thiele2014}, (ii) the opportunity to control the translational motion and trap Rydberg helium atoms using inhomogeneous electric fields~\cite{allmendinger2013,lancuba2013,lancuba2014,lancuba2016,palmer2017}, and (iii) the possibility of efficiently implementing resonance enhanced two-color two-photon Rydberg state excitation schemes using readily available continuous wave (cw) diode lasers.

For triplet states with $\ell>0$, the fine structure that arises as a result of the spin-orbit interaction must be accounted for when precisely determining transition frequencies and transition dipole moments, or when calculating the Stark structure of low-$|M_J|$ sublevels. The fine structure must also be considered when characterizing dephasing and decoherence in microwave transitions between these states, and in implementing microwave dressing schemes to reduce the sensitivity of low-$\ell$ states to static or radio-frequency electric fields~\cite{jones2013}.

The fine structure of the $1\mathrm{s}n\mathrm{p}\,^3\mathrm{P}_J$ levels in helium is typically referred to in terms of $\nu_{0, 2}$ and $\nu_{1, 2}$ -- the frequencies associated with the intervals between the $J=0$ and $J=2$ levels, and the $J=1$ and $J=2$ levels, respectively. These intervals scale with $n^{-3}$, whereas the sensitivity of the corresponding levels to electric fields increases with $n^7$. Consequently, for high values of $n$ even small stray electric or magnetic fields can cause Stark or Zeeman shifts that are comparable to the fine structure. 

Here we report studies of single-photon $1\mathrm{s}n\mathrm{s}\,^3\mathrm{S}_1 \rightarrow 1\mathrm{s}n\mathrm{p}\,^3\mathrm{P}_J$ transitions, for values of $n$ between 34 and 51. From the measured spectra $\nu_{0, 2}$ and $\nu_{1, 2}$ were both determined for $n=34$ to~36. In the spectra recorded for values of $n$ between~45 and~51 the smaller $\nu_{1, 2}$ interval could not be resolved. In these cases the interval $\nu_{0,\overline{1 2}}\simeq\nu_{0, 2}$ was measured. This interval corresponds to the energy difference between the $J=0$ level and the average energy of the unresolved $J=1$ and $J=2$ levels weighted by their multiplicities, $2J+1$. 

This article is structured as follows: Sec.~\ref{sec:background} provides an overview of the calculations employed to determine the intervals between the $1\mathrm{s}n\mathrm{p}\, ^3\mathrm{P}_J$ levels and the effects of weak magnetic and electric fields on this fine structure. Sec.~\ref{sec:experiment} contains a description of the experimental apparatus, the techniques used to prepare, probe and detect the high Rydberg states, and the methods used to characterize and minimize stray electric and magnetic fields.  The results of the measurements of the fine-structure of the $1\mathrm{s}n\mathrm{p}\,^3\mathrm{P}_J$ Rydberg levels are presented in Sec.~\ref{sec:results}. In Sec.~\ref{sec:conclusions} a discussion of the systematic uncertainties in the measurements is provided and conclusions are drawn.

\section{\label{sec:background}Calculations}

\subsection{\label{ssec:H0}Rydberg energy level structure}

In the absence of external electric or magnetic fields the Hamiltonian, $\hat{H}_{0}$, associated with a single excited Rydberg electron in the helium atom is diagonal in an $|n\,L\,S\,J\,M_J\rangle$ basis, where $n$ is the principal quantum number, $L\equiv\ell$ is the total electron orbital angular momentum quantum number, $S$ is the total electron spin quantum number, $J$ is the total angular momentum quantum number, and $M_J$ is the projection of $\vec{J}$ onto the laboratory quantization axis. The binding energy, $W(n, \delta)$, of each eigenstate is given by
\begin{eqnarray}\label{eq:ion_en}
    \frac{W(n, \delta)}{hc} &=& -\frac{R_{\mathrm{He}}}{{\left(n - \delta \right)} ^2},
\end{eqnarray}
where $h$ is the Planck constant, $c$ is the speed of light in vacuum, $\delta$ is the quantum defect, and $R_{\mathrm{He}}$ is the Rydberg constant for helium corrected for the reduced mass, $\mu_{\mathrm{He}} = m_{\mathrm{e}} M_{\mathrm{He}^+} / (M_{\mathrm{He}^+} + m_{\mathrm{e}})$, where $m_{\mathrm{e}}$ is the electron mass and $M_{\mathrm{He}^+}$ is the mass of the He$^+$ ion core. The value of the quantum defect depends on the quantum numbers $n$, $L$, $S$, and $J$, and can be determined using the recursive Ritz expansion~\cite{Drake1994}
\begin{equation}\label{eq:Ritz}
    \delta(n) = c_0 + \frac{c_2}{\left(n - \delta\right)^2} +\frac{c_4}{\left(n - \delta\right)^4} + \dots \;,
\end{equation}
where the values of $c_i$ for each set of $L$, $S$, and $J$ are obtained by fitting Eq.~\ref{eq:Ritz} to precise calculations~\cite{Drake1999} or measurements~\cite{Lichten1991, Sansonetti1992} of the corresponding Rydberg series. 
Higher order corrections to Eq..~\ref{eq:ion_en} have been reported~\cite{Drake1999} but their contributions are below the spectral resolution of the experimental apparatus used here.

\subsection{\label{ssec:Zeeman}Magnetic field effects}
In the presence of a weak magnetic field, $\vec{B}=(0,0,B_z)$, the total Hamiltonian for the Rydberg system is
\begin{equation}\label{eq:H_zeeman}
    \hat{H} = \hat{H}_0 + \hat{H}_\mathrm{Z}.
\end{equation}
The first-order perturbation by the magnetic field is
\begin{align}\label{eq:H_z}
    \hat{H}_\mathrm{Z} &= \mu_{B}(g_L\hat{\vec{L}} + g_S \hat{\vec{S}}) \cdot \vec{B} \nonumber \\
      &= \mu_{B}g_LB_z\hat{L}_{z} + \mu_{B}g_{S} B_z \hat{S}_{z},
\end{align}
where $\mu_B$ is the Bohr magneton, $g_S \approx 2.002\,319$ is the electron spin $g$-factor, and $g_L = 1 - m_{\mathrm{e}} / M_{\mathrm{He}^+}$ is the electron orbital $g$-factor. To first order in $\alpha$, the matrix elements associated with $\hat{H}_\mathrm{Z}$ are~\cite{Yan1994}
\begin{widetext}
\begin{eqnarray}\label{eq:yan}
\langle n'\,L'\,S\,J'\,M_J'|\hat{H}_\mathrm{Z}|n\,L\,S\,J\,M_J\rangle &=& \delta_{L,L'}\,\delta_{S,S'}\,\delta_{M_J,M_J'}\,\mu_{\mathrm{B}}\,B_z\,(-1)^{1-M_J}\sqrt{6(2J+1)(2J'+1)}\left(\begin{array}{ccc}
J' & 1 & J\\
-M_J & 0 & M_J
\end{array}\right)\times\dots\nonumber\\\nonumber\\
&&\hspace*{1.0cm}
\left[ 
(-1)^{J+J'+L+S}\,g_L'\,\left\{\begin{array}{ccc}
L & J' & S\\
J & L & 1
\end{array}\right\}
+ (-1)^{L+S}\,g_S'\,\left\{\begin{array}{ccc}
J & J' & 1\\
S & S & L
\end{array}\right\}
\right] \;,
\end{eqnarray}
\end{widetext}
where the terms in curved (curly) brackets represent Wigner 3J (6J) symbols. As in Ref.~\cite{Yan1994},
\begin{eqnarray}
g_{L}' &=& \sqrt{\frac{(2L+1)L(L+1)}{6}}\,g_{L},
\end{eqnarray}
and
\begin{eqnarray}
g_S' &=& \sqrt{\frac{(2S+1)S(S+1)}{6}}\,g_{S}.
\end{eqnarray}

\begin{figure}[h]
    \centering %
    \includegraphics[width=0.45\textwidth]{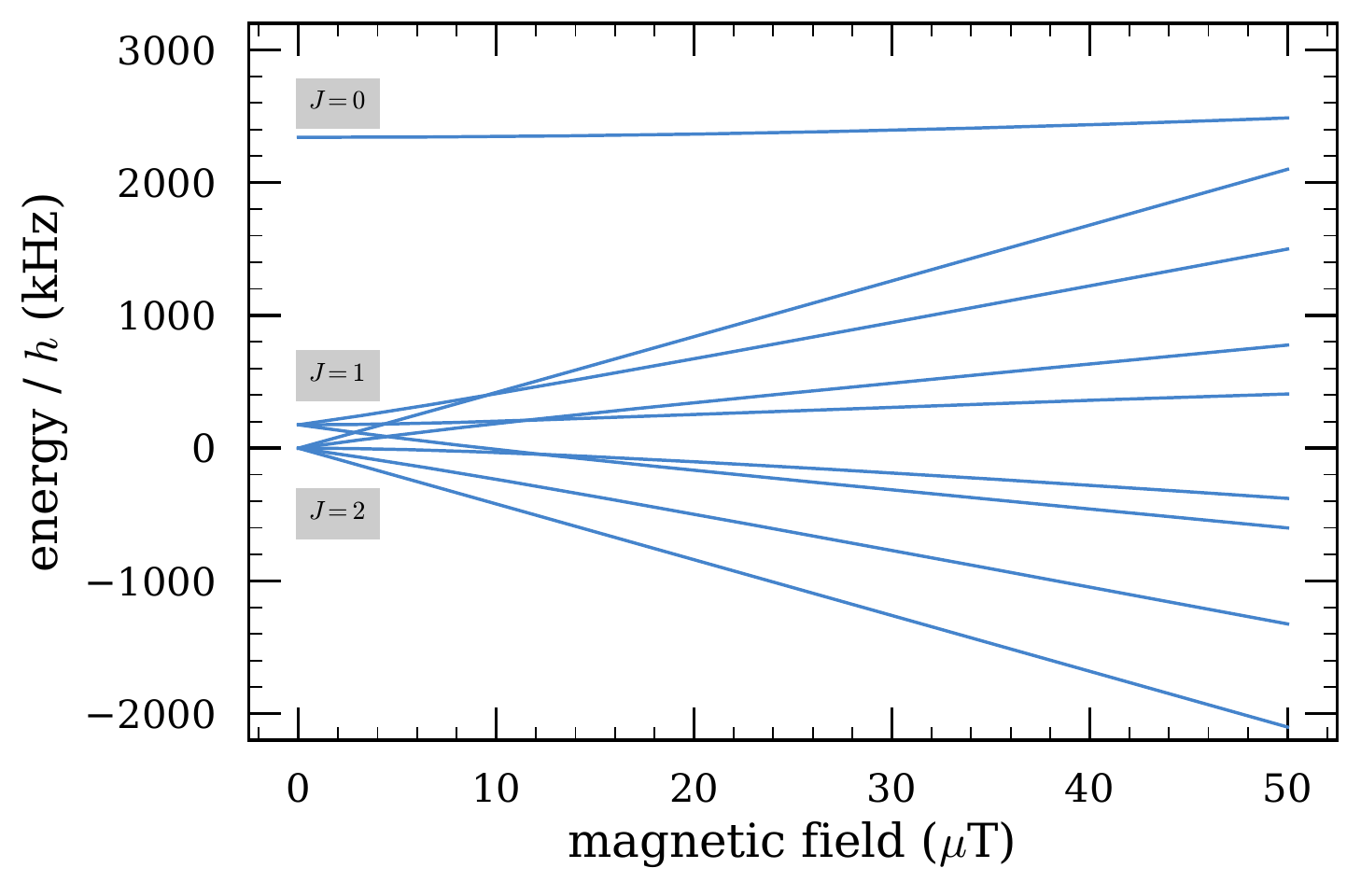} %
    \caption{\label{fig:zeeman_n45} Zeeman structure of the $1\mathrm{s}45\mathrm{p}\,^3\mathrm{P}$ term in helium. The vertical axis is displayed with respect to the energy of the $J=2$ level in zero magnetic field.}
\end{figure}

To calculate the energy level structure in the presence of a magnetic field the eigenvalues of the complete matrix containing the zero-field and Zeeman elements are then determined. The intensity of electric dipole transitions between pairs of Zeeman sublevels can be calculated by summing and squaring the contributions from each component of each $\langle n',L',S',J',M_J'|e\,\hat{\vec{r}}\,|n,L,S,J,M_J\rangle$ electric dipole transition moment, weighted by the corresponding elements of the initial and final state eigenvectors~\cite{edmonds1996}.

To evaluate the effects of weak magnetic fields on the $1\mathrm{s}n\mathrm{s}\,^3 \mathrm{S}_1$ and $1\mathrm{s}n\mathrm{p}\,^3 \mathrm{P}_J$ levels, the quantum defects in Ref.~\cite{Drake1999} were employed.  Figure~\ref{fig:zeeman_n45} shows the calculated energy level structure of the $1\mathrm{s}45\mathrm{p}\,^3 \mathrm{P}$ term for magnetic fields up to 50~$\mu$T. In this figure, both the large, $\nu_{0, 2}$, and small, $\nu_{1, 2}$, fine structure intervals can be identified in low fields.  As the magnetic field increases the smaller interval becomes obscured by the Zeeman splitting of the $1\mathrm{s}45\mathrm{p}\,^3\mathrm{P}_1$ and $1\mathrm{s}45\mathrm{p}\,^3\mathrm{P}_2$ levels. The $1\mathrm{s}45\mathrm{p}\, ^3\mathrm{P}_1$ ($M_J = -1$) and $1\mathrm{s}45\mathrm{p}\,^3\mathrm{P}_2$ ($M_J = + 2$) sublevels cross in a magnetic field of $B_z = 3$~$\mu$T.  Similar calculations indicate that the equivalent crossings occur at $6$~$\mu$T and 1.5~$\mu$T for $n=35$ and $n=55$, respectively.  This demonstrates the level to which stray magnetic fields must be reduced in order for the small fine-structure intervals to be resolved.

\begin{figure}[h]
    \centering %
    \includegraphics[width=0.45\textwidth]{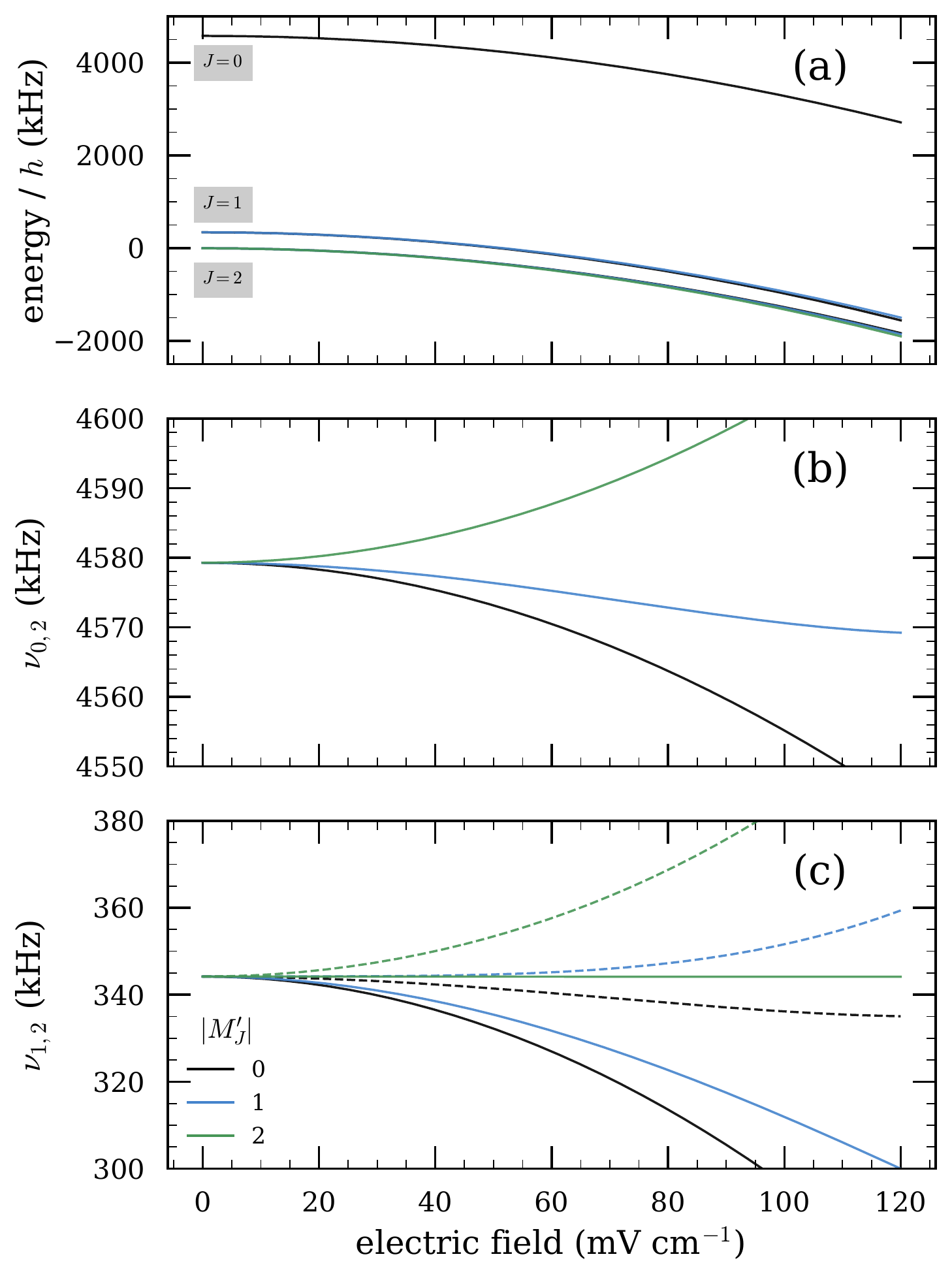} %
    \caption{\label{fig:stark_n36} (a) Calculated Stark structure of the $1\mathrm{s}36\mathrm{p}\,^3\mathrm{P}$ term in helium. The vertical axis is displayed with respect to the energy of the $J=2$ level in zero electric field. The calculated interval, $\nu_{0,2}$, between the $1\mathrm{s}36\mathrm{p}\,^3\mathrm{P}_0$ and $1\mathrm{s}36\mathrm{p}\,^3\mathrm{P}_2$ levels is shown in panel (b), while the smaller interval, $\nu_{1,2}$, between the $1\mathrm{s}36\mathrm{p}\,^3\mathrm{P}_1$ and $1\mathrm{s}36\mathrm{p}\,^3\mathrm{P}_2$ levels is shown in panel (c). The three curves in panel (b) represent the intervals between the $1\mathrm{s}36\mathrm{p}\,^3\mathrm{P}_0$ sublevel with $M_J = 0$ and the $1\mathrm{s}36\mathrm{p}\,^3\mathrm{P}_2$ sublevels with $|M_J| = 0, 1$ and~2. In panel (c) the three continuous (dashed) curves represent the intervals between the $1\mathrm{s}36\mathrm{p}\,^3\mathrm{P}_1$ sublevel with $M_J = 0$ ($|M_J| = 1$) and the $1\mathrm{s}36\mathrm{p}\,^3\mathrm{P}_2$ sublevels with $|M_J| = 0, 1$ and~2.}
\end{figure}

\subsection{\label{ssec:Stark}Electric field effects}
The effect that an electric field, $\vec{F} = (0, 0, F_z)$, has on the Rydberg levels is determined by calculating the eigenvalues of the Hamiltonian
\begin{equation}\label{eq:H_stark}
    \hat{H} = \hat{H_0} + \hat{H_\mathrm{S}},
\end{equation}
where,
\begin{equation}
    \hat{H_\mathrm{S}} = e F_z r \cos\theta.
\end{equation}
In the $|{n\,L\,S\,J\,M_J}\rangle$ basis the corresponding matrix elements can be expressed  as~\cite{Zimmerman1979}
\begin{widetext}
\begin{eqnarray}\label{eqn:stark}
\langle n'\,L'\,S'\,J'\,M_J'|\hat{H_\mathrm{S}}\,|n,L,S,J,M_J\rangle &=& \delta_{M_J,M_J'} e F_z\,\langle n'\,L'|r|n\,L\rangle\times\dots\nonumber\\
&&\hspace*{-5cm} \sum_{M_L=M_J+M_S} \left[(-1)^{L'+L-2S+2M_J}\sqrt{(2J+1)(2J'+1)}\left(\begin{array}{ccc}
L & S & J\\
M_L & M_J-M_L & -M_J
\end{array}\right)\times\dots\right.\nonumber\\
&&\hspace*{-3cm} \left.\left(\begin{array}{ccc}
L' & S & J'\\
M_L & M_J-M_L & -M_J
\end{array}\right)(-1)^{L'-M_L'}\sqrt{\max(L,L')}\left(\begin{array}{ccc}
L' & 1 & L\\
-M_L' & 0 & M_L
\end{array}\right)\right],
\end{eqnarray}
\end{widetext}
with the radial integrals, $\langle n'\,L'|r|n\,L\rangle$, calculated using the Numerov method \cite{Zimmerman1979}. For each value of $n$ of interest calculations were performed using three separate basis sets with $|M_J| = 0,\, 1$ or 2, and all of the corresponding Rydberg states within the range from $n-5$ to $n+5$.  A Stark map depicting the energy level structure of the $1\mathrm{s}36\mathrm{p}\,^3\mathrm{P}$ term in fields up to 120~mV\,cm$^{-1}$ is displayed in Fig.~\ref{fig:stark_n36}(a). For electric fields of $\sim50$~mV\,cm$^{-1}$ the fine-structure components all exhibit similar Stark shifts of $\sim320$~kHz. In fields of 10~mV\,cm$^{-1}$ this shift is 13~kHz and the corresponding changes in the intervals $\nu_{0, 2}$ and $\nu_{1, 2}$ between the levels are $\pm0.2$~kHz and $\pm0.4$~kHz, respectively. In the case of the $1\mathrm{s}51\mathrm{p}\,^3\mathrm{P}$ term, the Stark shifts of the individual levels in a field of 10~mV\,cm$^{-1}$ are $\sim250$~kHz and the changes in $\nu_{0, 2}$ and $\nu_{1, 2}$ are $\pm2$~kHz and $\pm5$~kHz, respectively.


\section{\label{sec:experiment}Experimental techniques}
\subsection{\label{ssec:rydberg}Rydberg helium spectroscopy}
\begin{figure}[h]
    \centering %
    \includegraphics[width=0.45\textwidth]{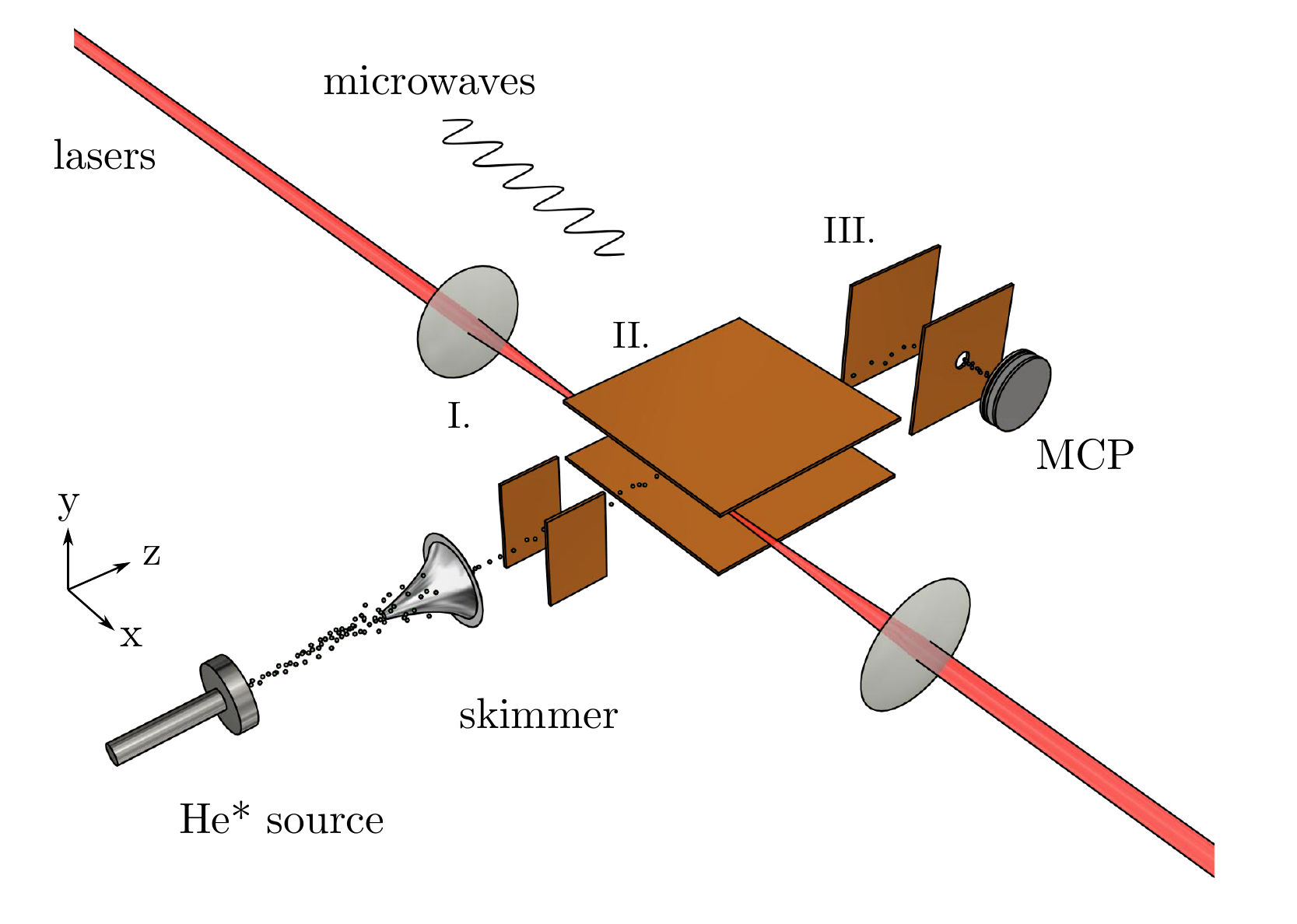} %
    \caption{\label{fig:schematic} Schematic diagram of the metastable helium beam source and Rydberg state photoexcitation and detection regions. Three pairs of parallel plate electrodes are used for (I.) ion deflection, (II.) to define the electric fields in the laser photoexcitation and microwave spectroscopy region, and (III.) to apply time-dependent electric fields to ionize the Rydberg atoms and accelerate the resulting electrons to an MCP detector.}
\end{figure}

A schematic diagram of the apparatus used in the experiments reported here is displayed in Fig.~\ref{fig:schematic}. Pulsed supersonic beams of helium atoms in the metastable $1\mathrm{s}2\mathrm{s}\,^3\mathrm{S}_1$ level were generated in an electric discharge at the exit of a pulsed valve operated at a repetition rate of 50~Hz. To maximise shot-to-shot stability, the discharge was seeded with electrons that emanated from a heated tungsten filament located $\sim 20$~mm downstream from the exit of the valve~\cite{Halfmann2000}. The beam passed through a 2-mm-diameter skimmer after-which stray ions produced in the discharge were removed by the electric field generated between a pair of parallel plate electrodes (region I. in Fig.~\ref{fig:schematic}).  The beam then proceeded to the photoexcitation region of the apparatus (region II. in Fig.~\ref{fig:schematic}), which was formed by a further two parallel 70 $\times$ 70~mm copper electrodes separated by 8.3~mm. Between these electrodes, co-propagating cw ultraviolet ($\lambda = 388.9751$~nm) and infrared ($\lambda \sim 788$~nm) laser beams were focused to intersect the collimated atomic beam. To excite a spatially localized bunch of atoms to Rydberg states, a pulsed electric field of $\sim 1$~V\,cm$^{-1}$ was applied in the excitation region for a period of $3$~$\mu$s.  The lasers were tuned to drive $1\mathrm{s}2\mathrm{s}\,^3\mathrm{S}_1 \rightarrow 1\mathrm{s}3\mathrm{p}\,^3\mathrm{P}_2 \rightarrow 1\mathrm{s}n\mathrm{s}\,^3\mathrm{S}_1$ transitions in this field.

After laser photoexcitation spectroscopy of single-photon $1\mathrm{s}n\mathrm{s}\,^3\mathrm{S}_1\rightarrow 1\mathrm{s}n\mathrm{p}\,^3\mathrm{P}_J$ transitions was performed using a pulsed source of microwave radiation.  The microwave pulses had a typical duration of 5~$\mu$s and entered the vacuum chamber 1~$\mu$s after the atoms were excited to the Rydberg levels. Population transfer between the Rydberg levels by the microwave radiation was identified by state-selective ramped-electric-field ionization between a third electrode pair (region III. in Fig.~\ref{fig:schematic}). The resulting electrons were then accelerated onto a microchannel plate (MCP) detector. The time-dependent ionization electric field was configured to ensure that the signals corresponding to field ionization of the initial and final Rydberg states were clearly distinguishable.

\subsection{\label{ssec:magnetic}Magnetic field control and characterization}

The magnetic field in the excitation region of the apparatus was controlled using three pairs of parallel coils wound directly onto the outside of the vacuum chamber and operated in a Helmholtz configuration. Each coil was 160~mm in diameter and the individual pairs were aligned with the $x$, $y$, or $z$ axes.

\begin{figure}[h]
    \centering %
    \includegraphics[width=0.45\textwidth]{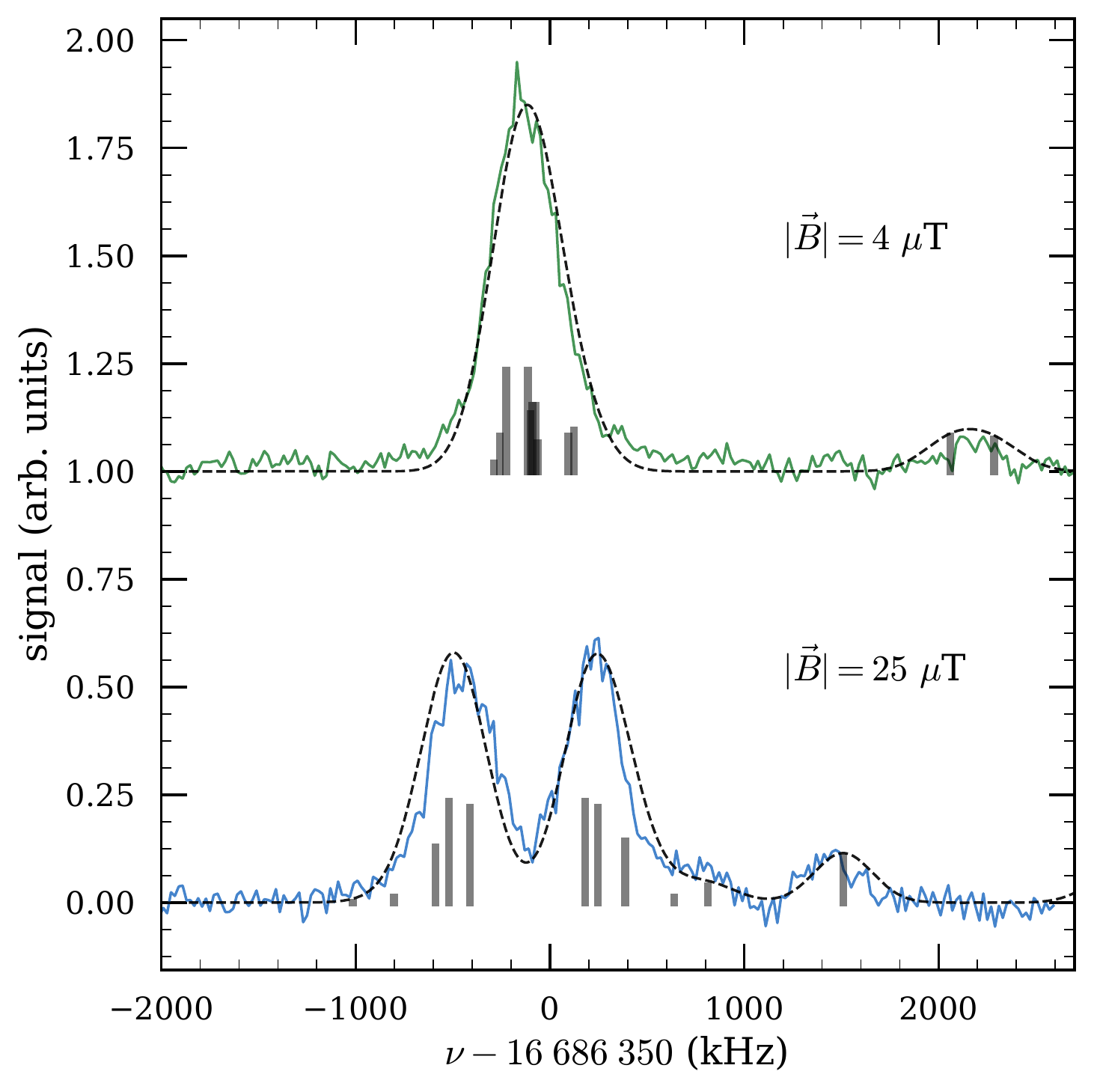} %
    \caption{\label{fig:45s-45p_mag} Microwave spectra of the $1\mathrm{s}45\mathrm{s}\,^3\mathrm{S}_1 \rightarrow 1\mathrm{s}45\mathrm{p}\,^3\mathrm{P}_{J}$ transitions measured in two different magnetic fields.  The vertical bars represent the set of allowed electric dipole transitions between the Zeeman-split sublevels and their relative intensities. The dashed curves are calculated spectra, shifted by $-170$~kHz (see text for details).}
\end{figure}

To minimize stray magnetic fields in region II (see Fig.~\ref{fig:schematic}), microwave spectra of the $1\mathrm{s}45\mathrm{s}\,^3 \mathrm{S}_1 \rightarrow 1\mathrm{s}45\mathrm{p}\,^3 \mathrm{P}_{J}$ transitions were recorded as the currents applied to the coils were systematically varied. Two examples of such spectra are shown in Fig.~\ref{fig:45s-45p_mag}.  For the minimum field configuration (upper spectrum in Fig.~\ref{fig:45s-45p_mag}), the 176~kHz fine-structure interval between the $J=1$ and $J=2$ levels of the $^3\mathrm{P}$ term could not be resolved. However, increasing or decreasing the currents gave rise to observable Zeeman splittings, as can be seen in the lower spectrum in Fig.~\ref{fig:45s-45p_mag}. The vertical bars overlaid with the recorded spectra represent the results of calculations of the relative intensities of single-photon electric-dipole transitions between the 3 Zeeman-split sublevels of the $^3\mathrm{S}_1$ level and the 9 Zeeman-split sublevels of the $^3\mathrm{P}$ term.  The dashed curves represent calculated spectra for magnetic fields of 4~$\mu$T and 25~$\mu$T, as indicated, which were obtained by convolving the corresponding sets of calculated transition frequencies and intensities with Gaussian functions with full-widths-at-half-maximum (FWHM) of $\Delta \nu = 350$~kHz.  The calculated transition frequencies  have been shifted by $-170$~kHz to bring them into line with the measured spectra.  This shift to the absolute $1\mathrm{s}45\mathrm{s}\,^3 \mathrm{S}_1 \rightarrow 1\mathrm{s}45\mathrm{p}\,^3 \mathrm{P}_{J}$ transition frequencies is attributed to a combination of effects, including Stark and Doppler shifts.  For reference, a stray electric field of 10~mV\,cm$^{-1}$ (see Sec.~\ref{ssec:electric}) is calculated to cause a shift in this transition frequency of 40~kHz. 

From the coil geometry and the currents applied to generate the cancellation magnetic field the magnitude of the stray field present when the coils were switched off was inferred to be $\sim 50$~$\mu$T. This is approximately equal to the magnitude of the Earth's magnetic field in London~(UK), at the time the experiments were performed, of $|\vec{B}| \approx 49$~$\mu$T~\cite{londonfield2016}. From the experimental resolution and calculated Zeeman shift, we estimate that this field was successfully cancelled in the measurement region to $\lesssim 4$~$\mu$T (compare the experimental data and the results of the calculations in the upper panel of Fig.~\ref{fig:45s-45p_mag}). 

\subsection{\label{ssec:electric}Electric field control and characterization}

\begin{figure}[h]
    \centering %
    \includegraphics[width=0.45\textwidth]{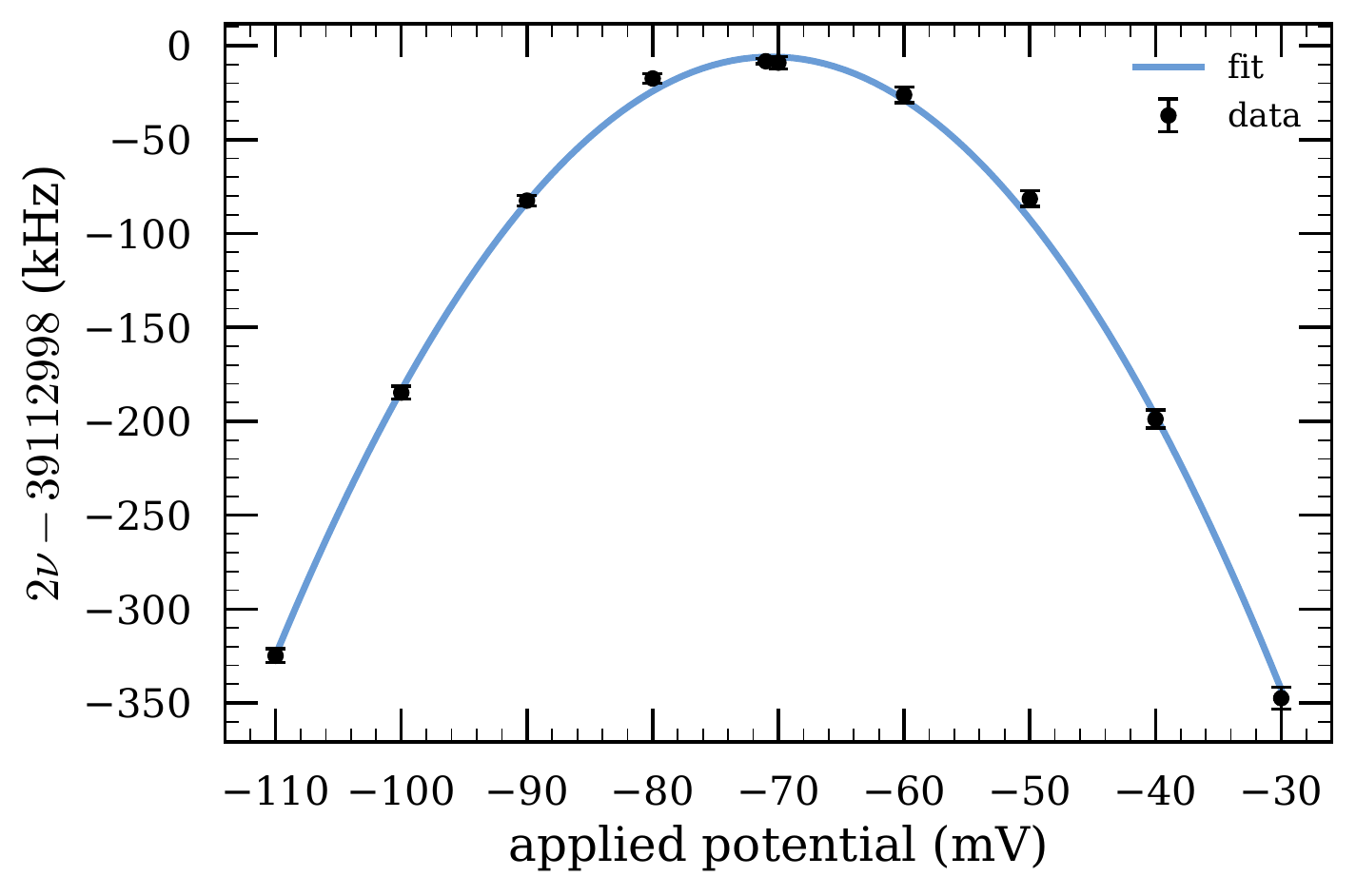} %
    \caption{\label{fig:55s-56s_2photon} The $1\mathrm{s}55\mathrm{s}\,^3\mathrm{S}_1 \rightarrow 1\mathrm{s}56\mathrm{s}\,^3\mathrm{S}_{1}$ two-photon transition frequency measured for a range of electric potentials applied to the upper electrode in the excitation region. The continuous curve represents a quadratic function fit to the data.}
\end{figure}

The two electrodes that demarcate region~II. in Fig.~\ref{fig:schematic} were used to control the electric field in the excitation region. However, these electrodes are themselves a source of stray electric fields, which arise from adsorbates, and patch and contact potentials~\cite{carter2011,neufeld2011}. To cancel contributions from these fields the electric potentials applied to the electrodes were adjusted to minimise the Stark shift of the two-photon $1\mathrm{s}55\mathrm{s}\,^3\mathrm{S}_1 \rightarrow 1\mathrm{s}56\mathrm{s}\,^3\mathrm{S}_1$ transition. The dependence of the measured transition frequencies, obtained by fitting Gaussian functions to the experimental data, on the offset potential applied to the upper electrode is displayed in Fig.~\ref{fig:55s-56s_2photon}. A quadratic function was fit to this data (continuous blue curve) to determine the optimal potential to minimize the stray electric field in the $y$ direction. In this process, the $1\mathrm{s}55\mathrm{s}\,^3\mathrm{S}_1 \rightarrow 1\mathrm{s}56\mathrm{s}\,^3\mathrm{S}_1$ transition frequency in the minimum achievable field was measured to be $2\nu = 39112992 \pm 2$~kHz.

The contributions to the stray electric field in the $x$ and $z$ directions cannot be cancelled with the electrode configuration employed. However, the magnitude of the residual uncancelled stray electric field could be determined from the difference between the minimum-field transition frequency and the zero-field transition frequency. 

\begin{figure}[h]
    \centering %
    \includegraphics[width=0.45\textwidth]{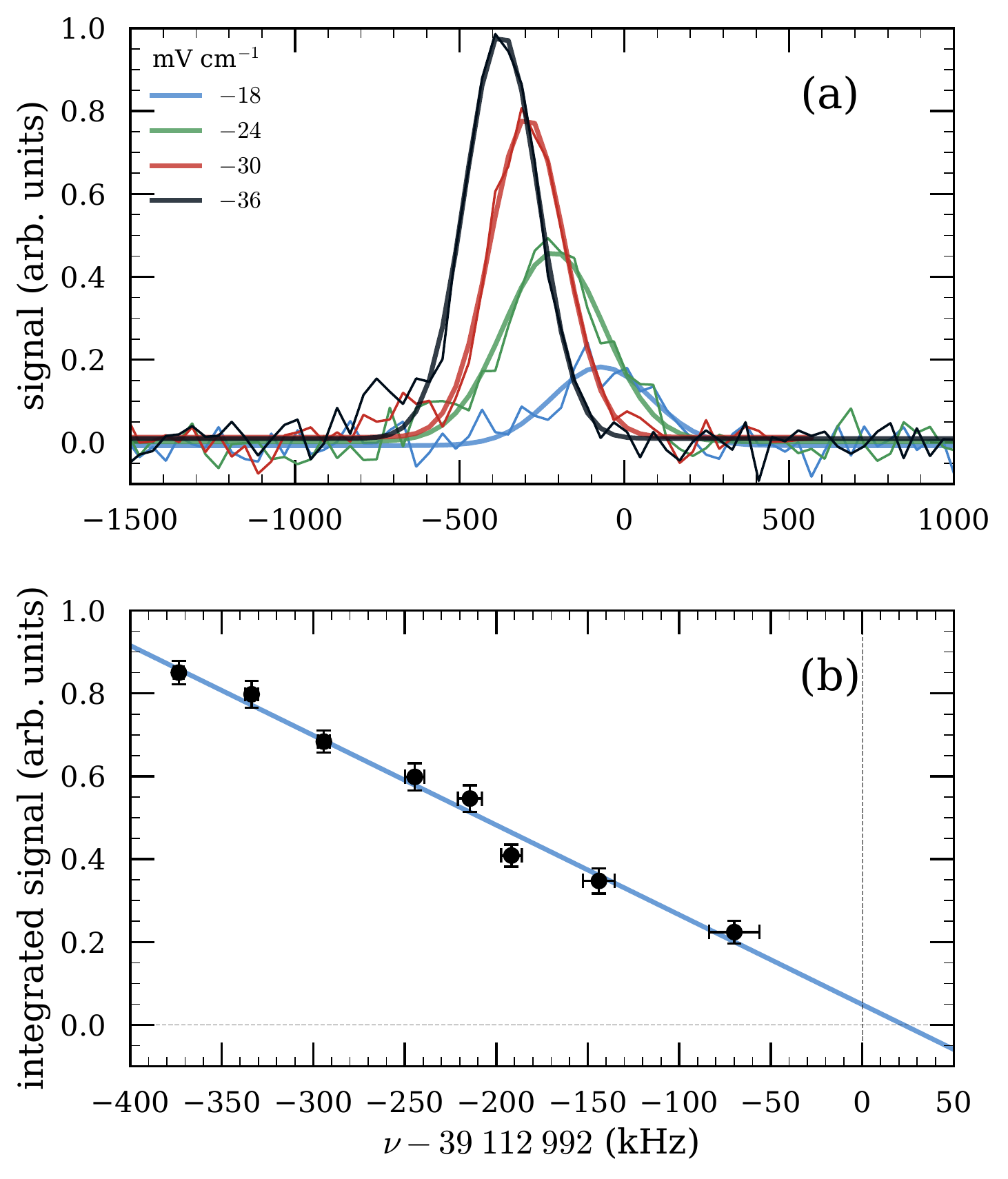} %
    \caption{\label{fig:55s-56s_efield} (a) Microwave spectra of the single-photon $1\mathrm{s}55\mathrm{s}\,^3\mathrm{S}_1 \rightarrow 1\mathrm{s}56\mathrm{s}\,^3\mathrm{S}_{1}$ transition in a range of electric fields.  The experimental data (thin curves) have been fit with Gaussian functions (thick curves). (b) Interdependence of the integrated spectral intensities and the transition frequencies in panel (a). The continuous line in panel (b) represents a least squares fit to the data.}
\end{figure}

To experimentally determine the zero-field $1\mathrm{s}55\mathrm{s}\,^3\mathrm{S}_1 \rightarrow 1\mathrm{s}56\mathrm{s}\,^3\mathrm{S}_1$ transition frequency the technique recently reported by Lee, Nunkaew and Gallagher~\cite{Lee2016} was employed. This involved recording single-photon microwave spectra of the $1\mathrm{s}55\mathrm{s}\,^3\mathrm{S}_1 \rightarrow 1\mathrm{s}56\mathrm{s}\,^3\mathrm{S}_1$ interval in a range of applied electric fields. As the fields were reduced toward zero, the intensity of this electric-dipole-forbidden single-photon transition reduces, as can be seen in Fig.~\ref{fig:55s-56s_efield}. In weak fields the spectral intensity of the transition decreases approximately linearly with the magnitude of the electric field. Therefore, the measured transition frequency can be extrapolated to zero intensity to obtain the zero-field transition frequency of $39113003 \pm 19$~kHz. The 11~kHz difference between this and the minimum-field transition frequency (measured by two-photon spectroscopy) suggests that the magnitude of the residual electric field in the apparatus during the measurements was $9 \pm 7$~mV\,cm$^{-1}$.  This estimate was made using the measured relative static electric dipole polarizability of the $1\mathrm{s}55\mathrm{s}\,^3\mathrm{S}_1$ and $1\mathrm{s}56\mathrm{s}\,^3\mathrm{S}_1$ states, as determined from the data in Fig.~\ref{fig:55s-56s_2photon}.

The $1\mathrm{s}55\mathrm{s}\,^3\mathrm{S}_1 \rightarrow 1\mathrm{s}56\mathrm{s}\,^3\mathrm{S}_1$ transition frequency calculated using the quantum defects in Ref~\cite{Drake1999} is $\nu_\mathrm{calc.} = 39112998.17 \pm 0.04$~kHz. The 6~kHz difference between this and the measured two-photon transition frequency implies that the residual uncancelled stray electric field was $7 \pm 1$~mV\,cm$^{-1}$. This estimate of the stray field obtained by comparison of the experimental data with the theoretical predictions is in agreement with that determined purely by experimental means.

\section{\label{sec:results}Results}
Having minimised the stray electric and magnetic fields in the experimental apparatus and determined the magnitude of the residual fields, the fine structure of the $1\mathrm{s}n\mathrm{p}\,^3\mathrm{P}_J$ levels was measured for values of $n$ from 34 to 36, and from 45 to 51. For each of these measurements $1\mathrm{s}n\mathrm{s}\,^3\mathrm{S}_1\rightarrow 1\mathrm{s}n\mathrm{p}\,^3\mathrm{P}_J$ transitions were driven with 5-$\mu$s-long pulses of microwave radiation. To ensure that AC Stark shifts and effects of saturation were minimal, the microwave intensity was set so that $<10\%$ population transfer occurred when the microwave radiation was resonant with the $1\mathrm{s}n\mathrm{s}\,^3\mathrm{S}_1\rightarrow 1\mathrm{s}n\mathrm{p}\,^3\mathrm{P}_2$ transition.  The spectra measured for the lower range of values of $n$ are presented in Fig.~\ref{fig:fine-structure_0,1-2}. Each spectrum is displayed with respect to the theoretically predicted $1\mathrm{s}n\mathrm{s}\,^3\mathrm{S}_1\rightarrow 1\mathrm{s}n\mathrm{p}\,^3\mathrm{P}_2$ transition frequency, $\nu_2$. In each case, the transition to the $J=0$ level is separated by approximately 5~MHz from the stronger transitions to the $J=1$ and $J=2$ levels.  Effects arising from the similarity of the fluorescence lifetime of these states ($\sim50~\mu$s at $n=35$) and the flight time of the atoms from region II. to region III. of the apparatus ($\sim60~\mu$s) caused a reduction in the spectral intensities of the transitions to the $1\mathrm{s}n\mathrm{p}\,^3\mathrm{P}_1$ and $1\mathrm{s}n\mathrm{p}\,^3\mathrm{P}_0$ levels compared to the spectral intensity of the transition to the $1\mathrm{s}n\mathrm{p}\,^3\mathrm{P}_2$ level for the lower values of $n$. Consequently, it was necessary to increase the microwave intensity by a factor of 2.5 (partial spectrum offset from the background level at the bottom of Fig.~\ref{fig:fine-structure_0,1-2}) to unambiguously identify the $1\mathrm{s}34\mathrm{s}\,^3\mathrm{S}_1\rightarrow 1\mathrm{s}34\mathrm{p}\,^3\mathrm{P}_0$ transition. The calculated frequencies and relative intensities of the transitions to the $J = 0$, 1, and 2 levels are represented by the vertical grey bars.  The measured values of $\nu_{0,2}$ and $\nu_{1,2}$ were extracted by fitting the sum of three independent Gaussian functions to each spectrum.  The results are presented in Table~\ref{tbl:FS} alongside the theoretically predicted fine-structure intervals.

\begin{figure}[h]
   \includegraphics[width=0.45\textwidth]{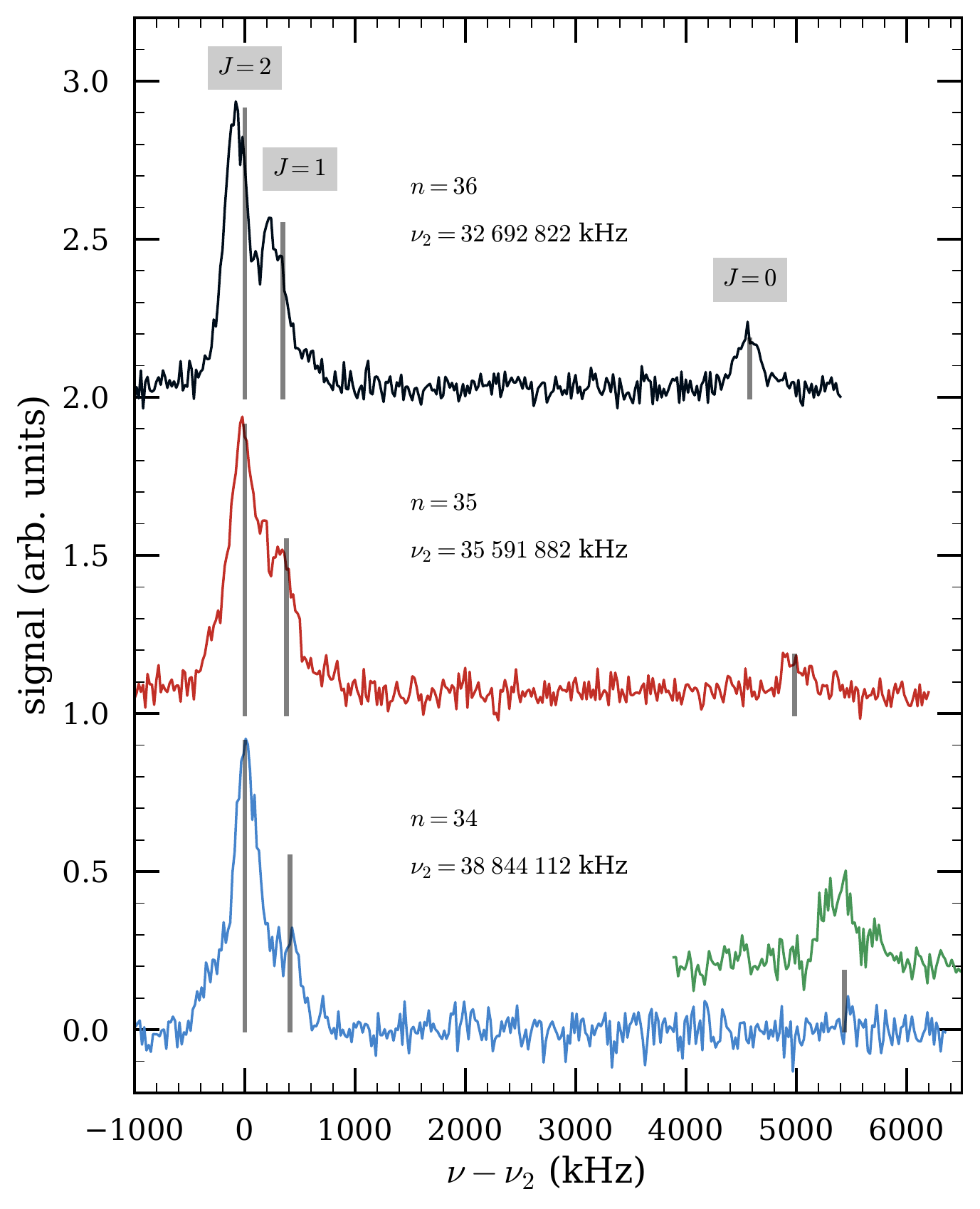} %
    \caption{\label{fig:fine-structure_0,1-2} Microwave spectra of the $1\mathrm{s}n\mathrm{s}\,^3\mathrm{S}_1 \rightarrow 1\mathrm{s}n\mathrm{p}\,^3\mathrm{P}_J$ transitions for $n=34$, 35 and 36. The frequency offset on the horizontal axis corresponds to the calculated $1\mathrm{s}n\mathrm{s}\,^3\mathrm{S}_1 \rightarrow 1\mathrm{s}n\mathrm{p}\,^3\mathrm{P}_2$ transition frequency, $\nu_2$. The calculated frequencies and relative intensities of the transitions to the $J = 0$, 1, and 2 levels are indicated by the vertical grey bars.  The partial spectrum, slightly offset from the background for $n=34$, was recorded with a microwave intensity 2.5 times higher than that used in recording the full spectrum.}
\end{figure}

\begin{table}[h]
\caption{\label{tbl:FS} Fine structure intervals between the $1\mathrm{s}n\mathrm{p}\,^3\mathrm{P}_J$ levels in helium.  The differences between the measured and calculated values, $\Delta \nu$, are given as a fraction of the combined uncertainty, $\sigma$, associated with fitting each pair of spectral features.}
\begin{ruledtabular}
\begin{tabular}{lrddd} \noalign{\vskip 1.5mm} %
& $n$ & \multicolumn{1}{c}{calc. (kHz)} & \multicolumn{1}{c}{exp. (kHz)} & \multicolumn{1}{c}{$\Delta\nu / \sigma$} \\ \noalign{\vskip 1.5mm}
 \colrule \\ 
$\nu_{1,2}$ & 34 & 408.73 & 397 & 0.9\\
           & 35 & 374.61 & 366 & 1.1\\
           & 36 & 344.19 & 340 & 0.8\\
\\
$\nu_{0,2}$  & 34 & 5437.87 & 5387 & 4.2\\
           & 35 & 4984.00 & 5014 & -1.2\\
           & 36 & 4579.27 & 4643 & -2.9\\
\\
$\nu_{0,\overline{12}}$%
           & 45 & 2275.66 & 2243 & 2.1\\
           & 46 & 2130.22 & 2150 & -0.9\\
           & 47 & 1996.92 & 1971 & 0.8\\
           & 48 & 1874.51 & 1817 & 2.8\\
           & 49 & 1761.90 & 1813 & -2.9\\
           & 50 & 1658.14 & 1646 & 0.4\\
           & 51 & 1562.36 & 1476 & 1.9\\
\end{tabular}
\end{ruledtabular}
\end{table}

In general, there is a good quantitative agreement between the experimentally measured and calculated fine-structure intervals for $n=34$ to 36. The fine-structure intervals decrease with $n^{-3}$ and it becomes more difficult to resolve $\nu_{1,2}$ for higher values of $n$.  This interval was not resolved for values of $n$ between 45 and 51 (see, e.g., upper spectrum in Fig.~\ref{fig:45s-45p_mag}). However, the $\nu_{0,\overline{1 2}}$ interval between the spectral intensity weighted average position of the $J=1$ and $J=2$ levels, and the $J=0$ level could be measured. These measured intervals together with the corresponding theoretical predictions are also included in Table.~\ref{tbl:FS}. 

The uncertainties stated in Table.~\ref{tbl:FS} are those associated with fitting the microwave spectra and do not include systematic contributions. Combining the results of the calculations in Sec.~\ref{ssec:Stark} with the measured residual uncancelled electric field of $\approx 10$~mV\,cm$^{-1}$ (see Sec.~\ref{ssec:electric}) allowed systematic uncertainties in the fine structure intervals caused by stray electric fields to be estimated. For fields $\sim10$~mV\,cm$^{-1}$, the Stark shifts of the $^3\mathrm{P}_J$ levels are approximately equal. Consequently, the fine-structure intervals at $n=36$ are shifted by $<1$~kHz (see Fig.~\ref{fig:stark_n36}). At $n=55$ this shift increases to $\sim10$~kHz. As can be seen from the data in Table.~\ref{tbl:FS}, frequency shifts on this scale are generally smaller than the error associated with fitting the experimental data.

The primary source of systematic uncertainty in the measured fine structure intervals originates from uncancelled static or time-dependent magnetic fields, which are estimated to have a magnitude $\lesssim 4$~$\mu$T. These fields can induce shifts of the $^3\mathrm{P}_J$ levels of up to $300$~kHz.  This is comparable to the $\nu_{1, 2}$ interval, and precludes the measurement of it for values of $n > 38$.

\begin{figure}[h]
    \centering %
    \includegraphics[width=0.45\textwidth]{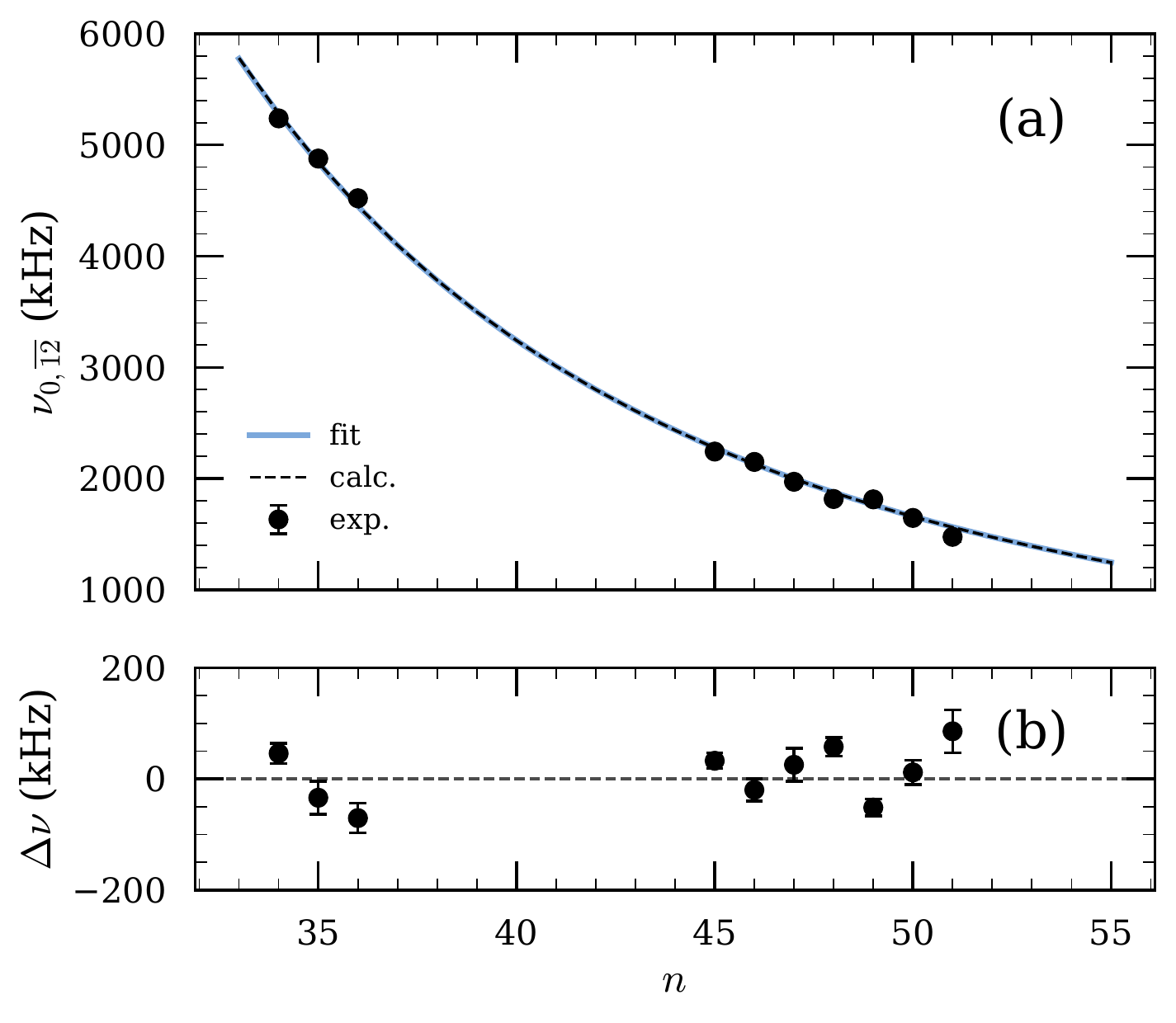} %
    \caption{\label{fig:fine-structure_n} (a) Measured values of the fine structure interval $\nu_{0,\overline{1 2}}$ for values of $n$ between~34 and~51 (points). The continuous blue curve represents the function $\nu = a n^{-3}$ fit to the experimental data such that $a = 2.076(9) \times 10^{8}$~kHz. The dashed black curve, which is indistinguishable from the fit, indicates the calculated values for the same interval. (b) The differences, $\Delta\nu$, between the measured and calculated data in panel (a).}
\end{figure}

Measured values of $\nu_{0,\overline{1 2}}$ are displayed in Fig.~\ref{fig:fine-structure_n})a= for all values of $n$ for which the experiments were performed. Over this range of states, the interval is observed to scale exactly with $n^{-3}$. This can be seen from the fit to the data in Fig.~\ref{fig:fine-structure_n}(a). The fit is indistinguishable from the dashed curve that was calculated using Eq.~\ref{eq:ion_en} as can be seen from the differences, $\Delta\nu$, between the measured and calculated data displayed in Fig.~\ref{fig:fine-structure_n}(b). The $n^{-3}$ dependence of the fine-structure intervals does not hold for low values of $n$ because of the increased complexity of the interaction of the Rydberg electron with the He$^+$ ion core. Indeed for values of $n\lesssim10$ for which the $1\mathrm{s}n\mathrm{s}\,^3\mathrm{P}_J$ fine structure has been precisely measured up to now, the exact $n^{-3}$ dependence is not valid.

\section{\label{sec:conclusions}Conclusions}

We have performed microwave spectroscopy of the $1\mathrm{s}n\mathrm{p}\,^3\mathrm{P}_J$ fine-structure in high Rydberg states of helium. The smaller, $\nu_{1, 2}$, and larger, $\nu_{0, 2}$, fine-structure intervals were both measured for $n=34, 35$, and 36.  For the more highly excited states studied in the range from $n=45$ to 51 only a doublet, corresponding to the $\nu_{0,\overline{1 2}}$ interval,  was resolved. For all of the states studied the measured intervals agree with the theoretical predictions obtained using the quantum defects reported in Ref.~\cite{Drake2002}. The typical full-widths-at-half-maximum of the measured spectral features associated with the $1\mathrm{s}n\mathrm{s}\,^3\mathrm{S}_1\rightarrow 1\mathrm{s}n\mathrm{p}\,^3\mathrm{P}_J$ transitions was $\sim 350$~kHz, and the uncertainty in determining the transition frequencies ranged from $\pm15$ to $\pm40$~kHz.

The measurement precision in the experiments was limited by the presence of stray electric and magnetic fields. The uncancelled electric field could be reduced in future measurements by careful preparation of the surfaces of the electrodes or the inclusion of additional electrodes to cancel the stray electric field in three dimensions~\cite{Witteborn1977, Dunning1993}.  Using this approach residual fields as low as 50~$\mu$V\,cm$^{-1}$ could be achieved. However, it will be necessary to carefully design the electrode structures to permit injection and unimpeded propagation of the microwave radiation. Presently, stray magnetic fields contribute more significantly to the experimental uncertainty than stray electric fields.  Calculations performed at $n=45$ indicate that a fifty-fold reduction in the magnitude of the stray magnetic field would be required to bring the uncontrolled Zeeman shifts down to the order of 1~kHz. This corresponds to fields of $|\vec{B}| \approx 100$~nT. A combination of magnetic shielding and active field cancellation could be used to realize magnetic fields $<10$~nT \cite{BITTER1991521}. If this were achieved then the spectral resolution of the microwave source and the finite interaction time of the atoms with the microwave field would set the limit on the overall uncertainty. In this case, colder beams of atoms and longer microwave pulses would be required to make further improvements.

The results presented here validate the theoretical predictions of the $1\mathrm{s}n\mathrm{p}\, ^3\mathrm{P}_J$ fine structure in Rydberg states of helium with high principal quantum numbers.  Improvements in the control of stray electric and magnetic fields in the experimental apparatus are expected to reduce the systematic uncertainties in the measured fine structure intervals to the level of 1~kHz.  The extreme sensitivity of high Rydberg states to stray electric and magnetic fields make it challenging to perform direct measurements that test bound state QED calculations.  However, as demonstrated, this sensitivity can be exploited for electrometry and magnetometry to allow accurate characterization of the fields. The methods used to accurately characterise these fields in the work reported here could also be implemented for field characterization and cancellation in precision measurements of the $1\mathrm{s}2\mathrm{p}\, ^3\mathrm{P}_J$ and $1\mathrm{s}3p\, ^3\mathrm{P}_J$ fine structure.  Our measurements of the Rydberg fine structure are of direct importance to hybrid cavity QED experiments involving helium atoms in triplet Rydberg states. They also provide important information which is required in the refinement of experiments to study resonant energy transfer in collisions of ammonia molecules with helium atoms in triplet Rydberg states~\cite{zhelyazkova2017}.

\begin{acknowledgments}
This work is supported by the Engineering and Physical Sciences Research Council under Grant No. EP/L019620/1, and the European Research Council (ERC) under the European Union's Horizon 2020 research and innovation program (grant agreement No 683341).
\end{acknowledgments}

\bibliography{lib}

\end{document}